\documentclass[preprint,12pt]{elsarticle}
\usepackage{amsmath,amssymb,amsfonts}
\usepackage{graphicx}
\usepackage{textcomp}
\usepackage{xcolor}
\usepackage[colorinlistoftodos]{todonotes}
\usepackage{float}
\usepackage{multirow} 

\usepackage[linesnumbered,ruled,vlined]{algorithm2e}

\def\BibTeX{{\rm B\kern-.05em{\sc i\kern-.025em b}\kern-.08em
    T\kern-.1667em\lower.7ex\hbox{E}\kern-.125emX}}

\usepackage{verbatim}       
\usepackage{hyperref}       

\usepackage{listings}

\definecolor{soliditykeyword}{rgb}{0.5, 0.0, 0.1}    
\definecolor{soliditystring}{rgb}{0.0, 0.3, 0.0}     
\definecolor{soliditycomment}{rgb}{0.3, 0.3, 0.3}    

\lstdefinelanguage{Solidity}{
    keywords={address, bool, string, uint, uint256, constructor, event, modifier, function, mapping, struct},
    keywordstyle=\color{soliditykeyword}\bfseries,
    sensitive=true,
    comment=[l]{//},
    morecomment=[s]{/*}{*/},
    commentstyle=\color{soliditycomment},
    stringstyle=\color{soliditystring},
}

\lstdefinestyle{solidity}{
    language=Solidity,
    basicstyle=\ttfamily\footnotesize, 
    numbers=left,
    numberstyle=\tiny\color{gray},
    breaklines=true,
    frame=single,
    rulecolor=\color{gray},
    showspaces=false,
    showstringspaces=false,
    tabsize=4,
}

\usepackage{algpseudocode}    

\RestyleAlgo{linesnumbered=false}

\journal{Blockchain: Research and Applications}

\begin{document}

\begin{frontmatter}



\title{Building Trust in Healthcare with Privacy Techniques: Blockchain in the Cloud} 


\author[stavanger]{Ferhat Ozgur Catak}
\author[stavanger]{Chunming Rong}
\author[stavanger]{Øyvind Meinich-Bache}
\author[laerdal]{Sara Brunner}
\author[stavanger]{Kjersti Engan}

\affiliation[stavanger]{organization={Department of Electrical Engineering and Computer Science, University of Stavanger},
            addressline={Kjell Arholms gate 41}, 
            city={Stavanger},
            postcode={4021}, 
            state={Rogaland},
            country={Norway}}

\affiliation[laerdal]{organization={Laerdal Medical},
            addressline={Tanke Svilands gate 30}, 
            city={Stavanger},
            postcode={4007}, 
            state={Rogaland},
            country={Norway}}

\begin{abstract}
This study introduces a cutting-edge architecture developed for the NewbornTime project, which uses advanced AI to analyze video data at birth and during newborn resuscitation, with the aim of improving newborn care. The proposed architecture addresses the crucial issues of patient consent, data security, and investing trust in healthcare by integrating Ethereum blockchain with cloud computing. Our blockchain-based consent application simplifies patient consent's secure and transparent management. We explain the smart contract mechanisms and privacy measures employed, ensuring data protection while permitting controlled data sharing among authorized parties. This work demonstrates the potential of combining blockchain and cloud technologies in healthcare, emphasizing their role in maintaining data integrity, with implications for computer science and healthcare innovation.
\end{abstract}

\begin{graphicalabstract}
\includegraphics[width=1.0\linewidth]{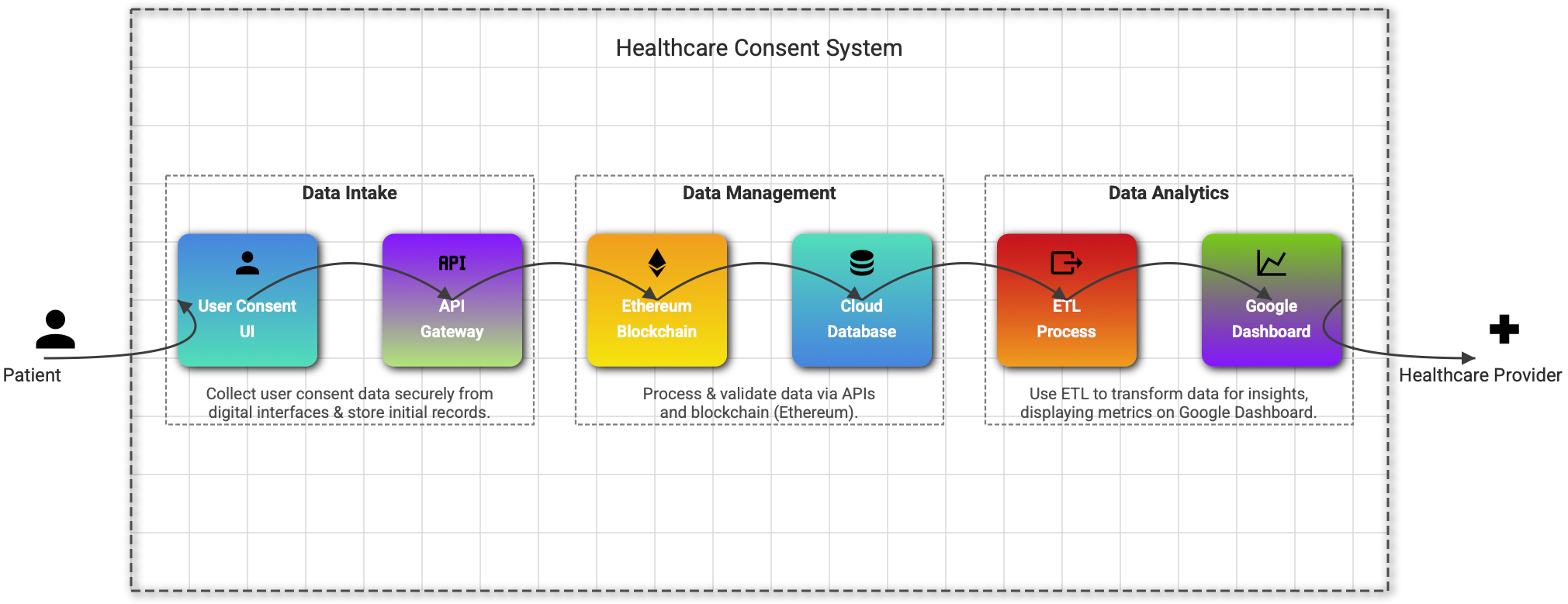}
\end{graphicalabstract}

\begin{highlights}
\item \textbf{Blockchain-Driven Consent Management:}
Developed a novel digital consent platform that leverages Ethereum blockchain and smart contracts to securely record, manage, and audit patient consent in healthcare. This solution ensures immutability, transparency, and tamper-proof recordkeeping.

\item \textbf{Enhanced Privacy and Data Security:}
Integrated advanced encryption techniques and data minimization principles to protect sensitive patient information while complying with privacy regulations such as GDPR. The decentralized architecture reduces the risk associated with centralized data breaches.

\item \textbf{Cloud-Integrated Architecture:}
Combined cloud computing with blockchain technology to enable a scalable infrastructure. The architecture supports secure video data collection and real-time consent verification, vital for projects like NewbornTime that aim to improve newborn care.

\item \textbf{User-Friendly Analytics Dashboard:}
Designed an intuitive, feature-rich dashboard that allows healthcare providers and researchers to monitor consent trends and access actionable insights. This enhances the decision-making process and bridges the gap between complex technologies and everyday clinical practice.

\item \textbf{Performance and Scalability Evaluation:}
Conducted extensive experimental analysis of gas consumption, transaction throughput, and scalability. The insights gained provide a solid foundation for further optimization and larger-scale deployment in healthcare settings.
\end{highlights}


\end{frontmatter}



\section{Introduction}

The collection of health data is of invaluable importance in medical-related research. While some data is routinely collected and stored, research projects often require the collection and storage of additional data. Informed consent is the general rule when collecting personal data for research \cite{griggs2018healthcare}. Traditionally, this has been done by signing a paper, which is then stored, and manually assigning a study ID to pseudonymize the data. This process is heavy, and a modern solution would include the digital collection of consent, sometimes called eConsent \cite{shah2024informed,obaidi2024econsent}. 

The recent European Health Data Space (EHDS) initiative exemplifies the future expectations for sharing health data. For data requiring consent, digital consent through a smart contract is a valuable solution \cite{zheng2020smartcontracts}, facilitating the giving, changing, and withdrawing of consent throughout the data's lifespan. For such digital consent systems to be widely adopted, they must be trustworthy. Users should be able to trust that personal information, such as names and identification numbers, and the key linking personal identification numbers to study IDs, are not seen or shared. Concerns about unauthorized access, data breaches, and privacy have increased the need for new solutions that protect patient rights and health data.

In this study, we propose a blockchain-based digital consent solution for research data, to be used within the \emph{NewbornTime}\footnote{\url{https://www.uis.no/en/research/newborntime}} project \cite{engan2023newborntime}. To ensure trustworthiness, all personal information is encrypted, and the correct consent information is kept together with all consent changes throughout the data's lifespan. The consent information is stored on a blockchain, ensuring tamper-proof consent records.

Globally, 10\% of newborns require assistance to start breathing after birth, and approximately 5\% need ventilation \cite{bjorland2019incidence, wyckoff2020neonatal}. \emph{NewbornTime – Improved Newborn Care based on video and artificial intelligence} is a research project that collects thermal video data from the time during labor, right before and after birth, as well as visual light video from the resuscitation tables for newborns needing assistance to start breathing. Using AI for activity recognition, the project aims to create objective timelines of events, such as birth and resuscitation interventions \cite{garcia2022towards, garcia2024comparative}. These timelines can be compared with guidelines and outcomes, supporting knowledge generation, debriefing, and quality improvement \cite{kolstad2024detection}.

Collecting and processing sensitive video data raises significant ethical and legal challenges. The NewbornTime project addresses these concerns through data minimization and a secure, multi-layered approach to ethical data handling using a secure cloud platform for video data storage. Strict access control and security protocols ensure compliance with the General Data Protection Regulation (GDPR). To manage patient consent, the project employs a blockchain-based system developed with BitUnitor\footnote{\url{https://www.bitunitor.com}}. This approach ensures transparency, immutability, and traceability of consent records, accessible only to authorized parties.

The main contributions of this study highlight significant advancements at the intersection of blockchain technology, cloud computing, and healthcare consent management.

\begin{enumerate}
    \item \textbf{Blockchain-Based Consent Application}: We present a blockchain-driven consent application specifically tailored for healthcare needs. This application securely and transparently manages patient consent, leveraging blockchain's tamper-proof and immutable features within the healthcare ecosystem.
    
    \item \textbf{Smart Contract Integration}: A central component of our solution is the use of smart contracts, which ensure secure, decentralized, and automated handling of consent transactions. By employing Ethereum's robust platform, the consent records remain accurate and resilient to unauthorized alterations.
    
    \item \textbf{Privacy and Security Measures}: Recognizing the sensitivity of healthcare data, our system incorporates advanced encryption techniques to safeguard personal information. Additionally, data minimization and decentralized storage ensure compliance with privacy regulations like GDPR, reducing vulnerabilities associated with centralized systems.
    
    \item \textbf{User-Friendly Dashboard for Consent Management}: To enhance accessibility and usability, we have integrated a consent management dashboard. This dashboard allows users to view, edit, and withdraw consents while providing insightful visualizations of consent trends over time. Its intuitive interface bridges the gap between technical innovations and user experience.
    
    \item \textbf{Broader Implications for Healthcare}: Beyond the NewbornTime project, our research demonstrates the transformative potential of combining blockchain and cloud technologies in healthcare. These innovations not only address current challenges in consent management but also pave the way for broader adoption of secure and transparent data-sharing practices across the sector.
\end{enumerate}

Our study represents a crucial step toward addressing the challenges of patient consent, and data security in healthcare. By presenting an integrated solution that combines blockchain, cloud computing, and a user-friendly dashboard within the scope of the NewbornTime project, we demonstrate the system's ability to ensure data integrity, and advance healthcare innovation. These contributions provide a strong foundation for future research and real-world implementation in diverse healthcare settings.

\section{Preliminary Information}

Before exploring the details of our architectural framework and its use in the NewbornTime project, we need to build a basic understanding of some key concepts that support our research. These concepts, such as blockchain, smart contracts, and Ethereum, provide the foundation for our approach.

\subsection{Blockchain}

Originally developed as the distributed ledger technology behind cryptocurrencies like Bitcoin, blockchain is a decentralized and unchangeable digital ledger represented as $\mathcal{L}$. It records transactions, where each transaction $T_i$ is a set $\langle S_i, R_i, D_i \rangle$ that includes a sender's address $S_i$, a recipient's address $R_i$, and data $D_i$. The ledger $\mathcal{L}$ is made up of blocks $\mathcal{B}_j$, so $\mathcal{L} = [\mathcal{B}_1, \mathcal{B}_2, \ldots, \mathcal{B}_n]$. Each block $\mathcal{B}_j$ holds a group of transactions and is linked to the previous block using cryptography, ensuring tamper-resistance through a hash function $\mathcal{H}: \mathcal{H}(\mathcal{B}_j) = \mathcal{B}_j{\text{(prev hash)}}$. This transparency and security make blockchain highly valuable in industries that need openness, and data integrity.

\subsection{Smart Contracts}

Smart contracts are self-executing, predictable programs represented as $\mathcal{C}$ and deployed on blockchain platforms like Ethereum. A smart contract $\mathcal{C}$ contains logic $\mathcal{L_C}$ as code, which sets conditions $\mathcal{C_{\text{conditions}}}$ that, when met, trigger actions $\mathcal{C_{\text{actions}}}$. In a pseudocode representation:

\begin{verbatim}
Smart Contract C:
    Conditions: C_conditions
    Actions: C_actions
\end{verbatim}

Smart contracts remove the need for intermediaries by automating contract execution based on predefined rules, promoting transparency, security, and efficiency. They are suitable for various uses, such as legal agreements, supply chain management, and managing healthcare consent.

\subsection{Ethereum}

Ethereum, represented as $\mathcal{E}$, is a leading blockchain platform that enhances the capabilities of blockchain technology \cite{szabo1997formalizing,buterin2014next}. It maintains a global state $\mathcal{S}$, made up of accounts $\mathcal{A}$, where each account $\mathcal{A}_i$ is identified by an address $\mathcal{A}_i{\text{address}}$. Ethereum allows developers to create and deploy smart contracts, enabling a wide range of decentralized applications (DApps). Ethereum's Turing-complete scripting language and flexible ecosystem make it a great choice for projects that combine blockchain with innovative contract capabilities across various fields, including healthcare.

With this basic understanding of blockchain, smart contracts, and Ethereum, we are ready to dive into the details of our architectural framework and how it is used in the NewbornTime project. Our approach uses these technologies to tackle the important challenges of patient consent, data security, all supported by a strong theoretical foundation.

\section{Web Interface for Users}

The web interface of the NewbornTime project has been designed to provide a user-friendly platform for users to give, view, and manage their consent related to participation in the project. Below are the key components of the interface with relevant screenshots.

\subsection{User Consent Submission}

\begin{figure}[!htbp]
    \centering
    \includegraphics[width=0.9\linewidth]{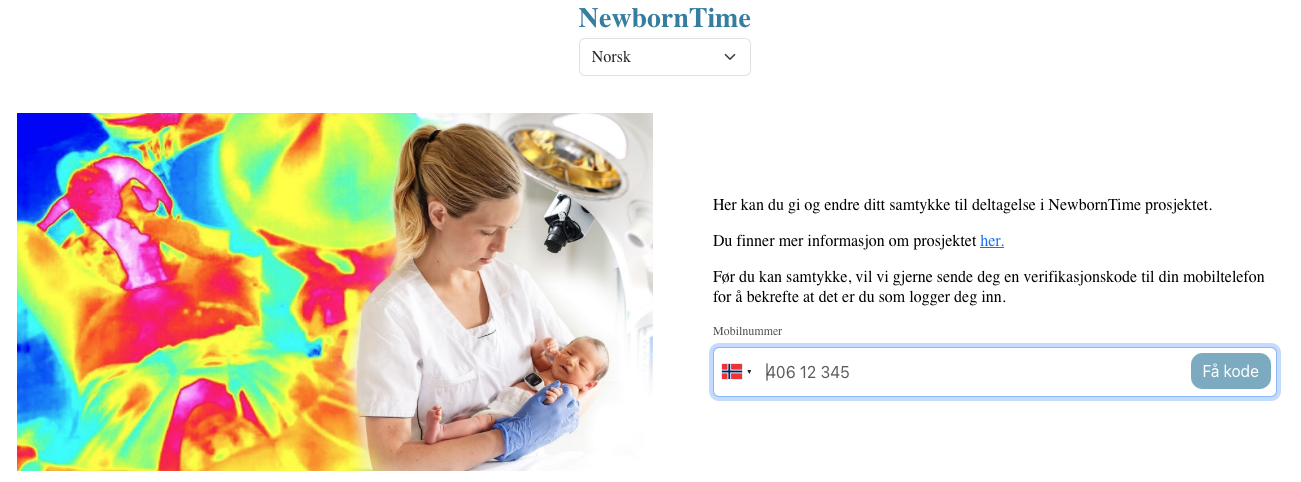}
    \caption{Initial screen for user consent submission, requiring phone number verification.}
    \label{fig:nbt1}
\end{figure}

When a user accesses the platform, they are prompted to provide their phone number to receive a verification code, ensuring that the user giving consent is verified (Figure~\ref{fig:nbt1}). This step is crucial for verifying that the correct individual is interacting with the system.

The system also supports the entry of paper-signed consents by research staff, ensuring that all consent records, whether digital or paper-based, are centralized and accessible for verification. This feature accommodates scenarios where paper signing is more practical, such as during initial participant recruitment.

\subsection{User Dashboard}

\begin{figure}[!htbp]
    \centering
    \includegraphics[width=0.5\linewidth]{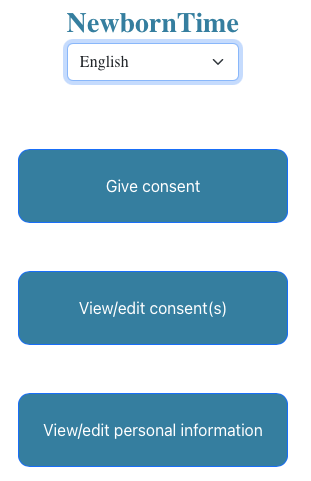}
    \caption{User dashboard offering options to give consent, view/edit consents, or modify personal information.}
    \label{fig:nbt2}
\end{figure}

After verification, users are presented with a dashboard where they can give consent, view or edit existing consents, and modify personal information (Figure~\ref{fig:nbt2}). The options are straightforward and designed to allow users to quickly navigate the platform.

\subsection{Consent Overview}

\begin{figure}[!htbp]
    \centering
    \includegraphics[width=0.6\linewidth]{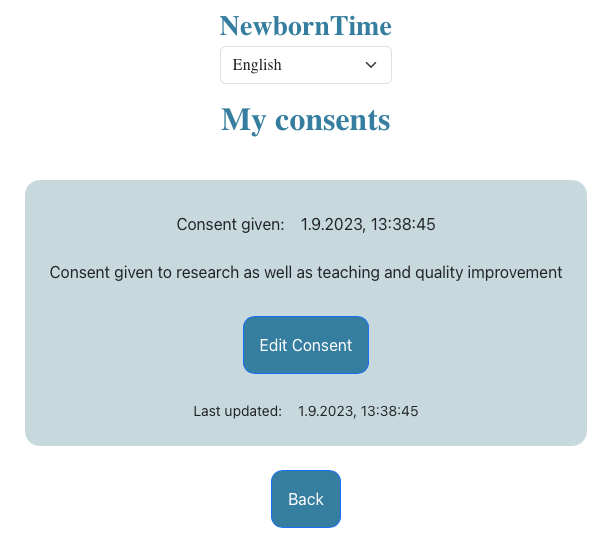}
    \caption{Overview of the user’s given consent with options to edit or withdraw consent.}
    \label{fig:nbt3}
\end{figure}

In the consent overview section (Figure~\ref{fig:nbt3}), users can view the details of their current consent, including the type of consent given and the time and date the consent was submitted. The interface allows the user to edit or withdraw consent as needed.

\subsection{Consent Editing}

\begin{figure}[!htbp]
    \centering
    \includegraphics[width=0.5\linewidth]{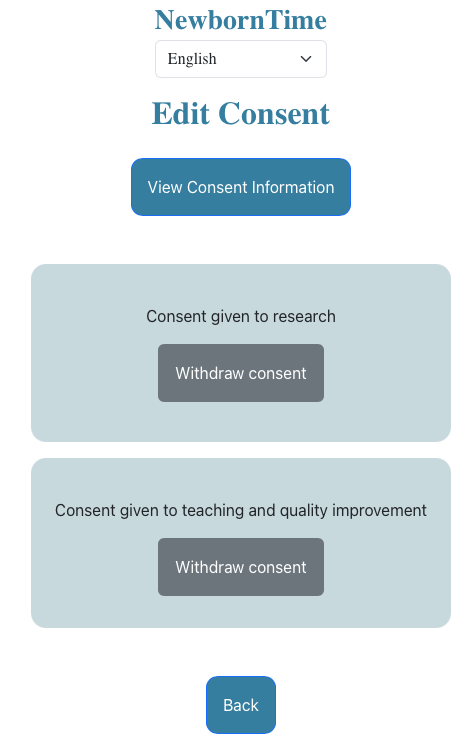}
    \caption{Consent editing interface where users can withdraw their consent for specific purposes.}
    \label{fig:nbt4}
\end{figure}

Users can withdraw specific consent through the consent editing interface (Figure~\ref{fig:nbt4}). This feature allows flexibility, giving users control over their consent at any time. The interface provides clear options to withdraw consent for different purposes, such as research or teaching.

The web interface has been developed with a focus on ease of use, security, and transparency, ensuring that users can efficiently manage their consent and personal information in line with the project's ethical standards.

\subsection{Consent Statistics Dashboard}

\begin{figure}[!htbp]
    \centering
    \includegraphics[width=1.0\linewidth]{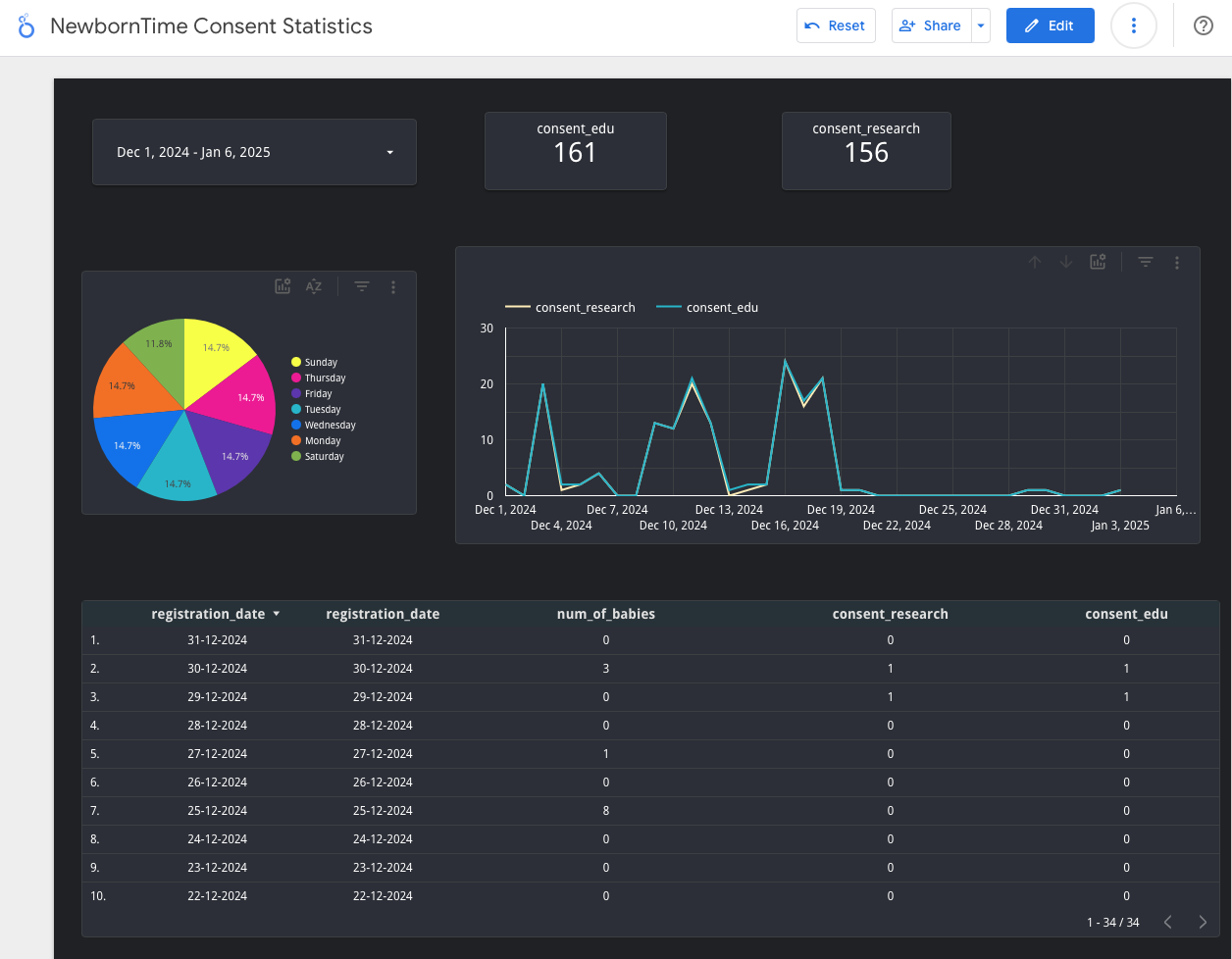}
    \caption{Consent Statistics Dashboard displaying trends, distributions, and detailed records for consent management.}
    \label{fig:nbt_dashboard}
\end{figure}

The Consent Statistics Dashboard is an integral feature of the NewbornTime web interface, designed to provide both administrators and researchers with valuable insights into consent trends over time. This interactive dashboard visualizes key metrics, allowing stakeholders to monitor and analyze consent data effectively (Figure~\ref{fig:nbt_dashboard}).

The dashboard includes the following features:
\begin{itemize}
    \item \textbf{Consent Trends Over Time}: A line graph displays the number of consents given for different purposes, such as research or education, over a selected time range. This feature helps to track consent activity and identify patterns.
    \item \textbf{Weekly Distribution of Consents}: A pie chart illustrates the distribution of consents across days of the week, providing insights into user behavior and engagement patterns.
    \item \textbf{Summary Metrics}: Key statistics, such as the total number of consents for education and research, appear to offer a quick overview of the current consent status.
    \item \textbf{Detailed Consent Records}: A tabular view lists individual consent records, including registration date, number of participants, and types of consent given. This granular data supports in-depth analysis and reporting.
\end{itemize}

\section{Consent Management Smart Contract}

This section provides a detailed look at the architecture behind the NewbornTime project, with a specific focus on the Ethereum-based healthcare consent management system. 

\subsection{System Overview}

\begin{figure}[!htbp]
    \centering
    \includegraphics[width=1.0\linewidth]{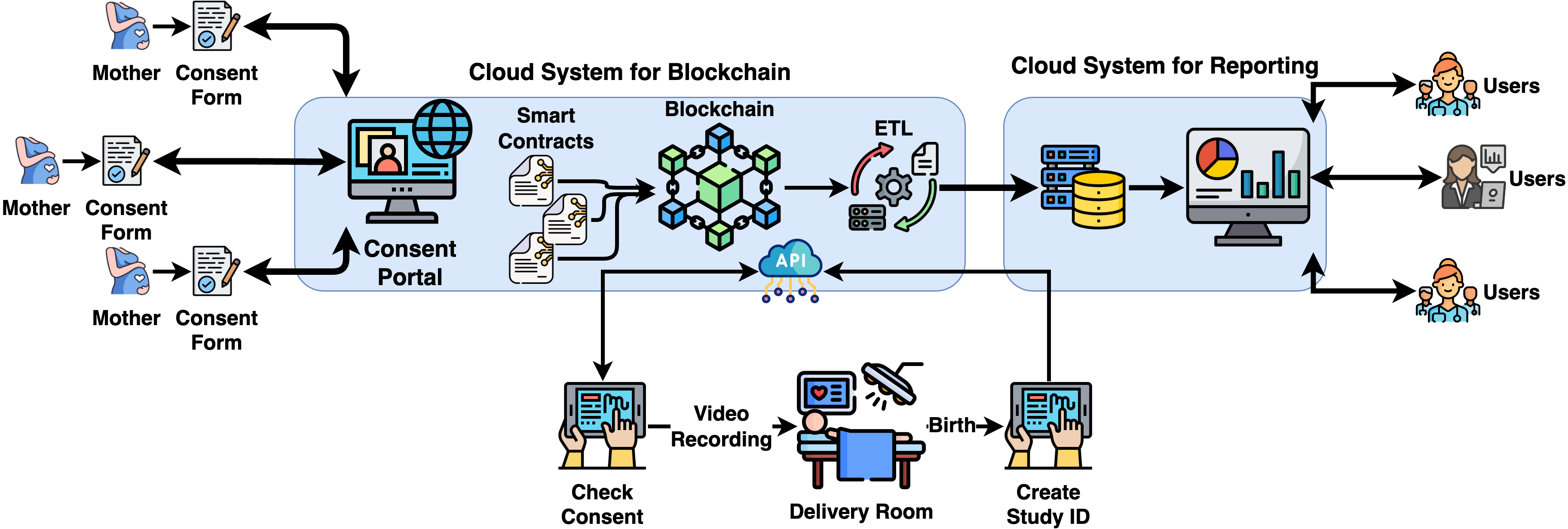}
    \caption{System overview of the NewbornTime consent management system, showcasing the integration of the consent portal, blockchain, cloud reporting, and real-time consent verification processes.}
    \label{fig:system_overview}
\end{figure}

Figure~\ref{fig:system_overview} illustrates the end-to-end architecture of the consent management system implemented for the NewbornTime project. This modular design seamlessly integrates blockchain technology, cloud-based infrastructure, and user-friendly interfaces to address critical challenges in healthcare consent management. The system is composed of the following main components:

\begin{itemize}
    \item \textbf{Consent Portal:} Mothers submit their consent through an intuitive web interface, which securely logs all actions. The portal also allows for modifications, withdrawals, and updates to consent records.
    \item \textbf{Cloud System for Blockchain:} Consent data is managed by smart contracts deployed on the Ethereum blockchain. These contracts ensure secure and immutable storage of consent records, handling all submissions, withdrawals, and updates while providing tamper-proof event logs.
    \item \textbf{Cloud System for Reporting:} This component includes an Extract-Transform-Load (ETL) process that aggregates and anonymizes data from the blockchain. The data is then visualized through an interactive dashboard, providing healthcare professionals and researchers with actionable insights into consent trends, compliance rates, and distribution patterns.
    \item \textbf{Consent Verification for Delivery Room:} The system ensures real-time verification of consent during critical moments such as video recordings in the delivery room. Healthcare staff can generate unique study IDs linked to the consent records, ensuring compliance and ethical data handling.
\end{itemize}

The consent management system is tailored for research purposes, facilitating the collection and management of consent for research data, distinct from clinical healthcare data. Currently, the system is implemented for the NewbornTime research project, managing consent for video data analysis during birth and resuscitation. Its architecture, however, allows for potential future applications in broader healthcare settings, pending additional regulatory and usability assessments \cite{albanese2020dynamic}. The system integrates both digital and paper-based consent processes, allowing for flexibility in how consents are obtained and recorded. The system integrates with the NewbornTime project by verifying consent before uploading video data to the cloud storage. The process involves: 1) Recording births and consents in the Liveborn Observation app, 2) Generating study IDs via the BitYoga system, 3) Periodically checking for valid consents and uploading corresponding videos to Azure if consent is active.

\subsection{Architectural Formalization}

The consent management system employs a robust, decentralized architecture designed to enhance newborn care by seamlessly integrating advanced AI algorithms for video analysis during childbirth and postpartum. The architecture also facilitates the secure and traceable management of patient consent, including the generation of unique Study IDs to link maternal and neonatal data.

The blockchain-based consent management algorithm presented in Algorithm~\ref{alg:consent} outlines the core processes involved in submitting, querying, creating Study IDs, and withdrawing consent on the Ethereum blockchain.

The \texttt{SubmitConsent} procedure is initiated when a mother accesses the consent management platform. The process begins with user verification, where a verification code is sent to the mother’s phone, and the user is required to submit this code to confirm their identity. If the verification is successful, the mother selects the type of consent they wish to provide (e.g., for research or education purposes). The consent record is encrypted using the mother’s public key to ensure confidentiality before being sent to the smart contract on the Ethereum blockchain. The smart contract stores the consent and emits a \texttt{ConsentGiven} event, signaling that the consent has been successfully recorded. At this stage, only a Mother ID is generated by the system, and no Baby ID is assigned yet.

The \texttt{QueryConsent} procedure enables to verify whether a mother has given consent. The provider (e.g. Laerdal Medical) uses the mother’s personal number as a key parameter to query the blockchain for the consent record. The smart contract retrieves the encrypted consent record. The consent information is made available to the provider. This step is critical during childbirth when video recordings or other data collection processes require immediate consent verification.

The \texttt{CreateStudyID} procedure is executed post-birth. If valid consent is confirmed for the mother, the research assistant can invoke an API call to generate a unique Study ID that links the Mother ID and the newly assigned Baby ID. The Study ID serves as a pseudonymized identifier for securely associating consent records with data collected during and after birth. The smart contract records the Study ID creation and emits a \texttt{StudyIDCreated} event for transparency.

The \texttt{WithdrawConsent} procedure allows the user to revoke specific consents through the platform. The smart contract updates the consent record on the blockchain to reflect withdrawal and emits a \texttt{ConsentWithdrawn} event, ensuring that the revoked consent is no longer valid for future use.

The algorithm ensures the security and privacy of user data through encryption and access control mechanisms. All consent records are encrypted before being sent to the blockchain, ensuring that only authorized entities can access and view the records. The smart contract acts as a gatekeeper, enforcing strict authorization protocols while maintaining transparency through event emissions.

\begin{algorithm}[h]
\footnotesize
\caption{\small Blockchain-Based Consent Management Algorithm}\label{alg:consent}
\SetNlSty{}{}{} 
\KwIn{User data (phone number, personal number), consent type}
\KwOut{Consent record stored on the blockchain, Study ID generated}
\textbf{SubmitConsent:}\\
    Mother accesses platform and submits verification code\\
    \eIf{verification successful}{
        Mother selects consent type\\
        Encrypt consent record and send to smart contract\\
        \textbf{Smart Contract:} Store consent, generate Mother ID, and emit \texttt{ConsentGiven} event
    }{
        Terminate process
    }
\textbf{QueryConsent:}\\
    Healthcare provider submits mother's personal number\\
    \eIf{authorized}{
        Retrieve and decrypt consent record\\
    }{
        Deny access
    }
\textbf{CreateStudyID:}\\
    Provider queries consent status using mother’s personal number\\
    \eIf{consent valid}{
        Generate unique Study ID by combining Mother ID and Baby ID\\
        \textbf{Smart Contract:} Record Study ID and emit \texttt{StudyIDCreated} event\\
    }{
        Deny Study ID creation
    }
\textbf{WithdrawConsent:}\\
    User selects consent to withdraw\\
    \textbf{Smart Contract:} Mark consent as withdrawn and emit \texttt{ConsentWithdrawn} event
\end{algorithm}

\subsection{Ethereum Smart Contract}

In the ensuing discussion, we present the Ethereum smart contract titled \texttt{HealthcareConsent.sol}. This contract occupies a pivotal role within the project, empowering the secure administration of patient consents via the Ethereum blockchain.

\begin{lstlisting}[style=solidity, caption={HealthcareConsent.sol Smart Contract}]
// SPDX-License-Identifier: MIT
pragma solidity ^0.8.0;

contract HealthcareConsent {
    address public owner;
\end{lstlisting}

Starting with a SPDX-License-Identifier comment, this contract sets the licensing terms under which the code runs, following the MIT license.

Named \texttt{HealthcareConsent}, this contract includes a state variable, \texttt{owner}, which is used to store the Ethereum address of the contract's owner.

\begin{lstlisting}[style=solidity]
    // Struct to represent a consent record
    struct Consent {
        address patient;
        address healthcareProvider;
        bool isConsentGiven;
        string motherName;
        uint256 nationalID;
        string phoneNumber;
        uint256 timestamp;
    }
\end{lstlisting}

- A \texttt{Consent} struct is introduced to hold individual consent records. This structure includes various attributes, such as the patient's Ethereum address, the healthcare provider's address, the consent status (\texttt{isConsentGiven}), the mother's name (\texttt{motherName}), the national identification number (\texttt{nationalID}), the phone number (\texttt{phoneNumber}), and a timestamp (\texttt{timestamp}).

\begin{lstlisting}[style=solidity]
    // Mapping from patient's address to their consents
    mapping(address => Consent[]) public patientConsents;
\end{lstlisting}

- A mapping named \texttt{patientConsents} is leveraged to aggregate arrays of \texttt{Consent} records, categorizing them under the Ethereum addresses of respective patients. In this structure, the Ethereum address of the patient serves as the mapping's key, while the value corresponds to an array of \texttt{Consent} records.

\begin{lstlisting}[style=solidity]
    // Event to log consent changes
    event ConsentChanged(address indexed patient, address indexed healthcareProvider, bool isConsentGiven, uint256 timestamp);
\end{lstlisting}

- The contract incorporates an \texttt{event} termed \texttt{ConsentChanged}, intended to record alterations in consent statuses. This event is distinguished by indexed parameters, namely \texttt{patient}, \texttt{healthcareProvider}, \texttt{isConsentGiven}, and \texttt{timestamp}, serving as pivotal transparency and audit trail mechanisms.

\begin{lstlisting}[style=solidity]
    constructor() {
        owner = msg.sender;
    }
\end{lstlisting}

- The contract's constructor initializes the \texttt{owner} variable, assigning it the Ethereum address of the contract deployer, represented by \texttt{msg.sender}. This address designates the contract's owner.

\begin{lstlisting}[style=solidity]
    // Modifier to restrict access to the contract owner
    modifier onlyOwner() {
        require(msg.sender == owner, "Only the contract owner can call this function");
        _;
    }
\end{lstlisting}

- The contract includes a \texttt{modifier} called \texttt{onlyOwner}, which is used to limit access to certain functions. Functions with this modifier can only be called by the contract owner. This access control works by checking that the sender's Ethereum address (\texttt{msg.sender}) matches the contract owner's address.

\subsection{Patient Consent Record}

Each consent record housed within the \texttt{HealthcareConsent.sol} contract is distinguished by the following attributes:

\begin{itemize}
    \item Patient's Ethereum address
    \item Educational and research consent status (either granted or revoked)
    \item Mother's name
    \item National identification number
    \item Phone number
    \item Timestamp
    \item Study ID
\end{itemize}

These attributes together enable the careful tracking and management of patient consents, ensuring their integrity and accessibility.

\subsection{Security Considerations}

\subsubsection{Access Control Mechanisms}

Access to important functions within the smart contract is controlled by the \texttt{onlyOwner} modifier, written as $\texttt{onlyOwner}()$. This modifier restricts access, allowing only the contract owner ($\texttt{msg.sender} = \texttt{owner}$) to execute these functions, usually a trusted entity in the healthcare system. Additionally, each patient is given a unique Ethereum address which improves identity verification.

\subsubsection{Auditability and Transparency}

To support auditing and maintain transparency, the system uses the \texttt{ConsentChanged} event. This event logs all changes in consent status, creating a clear and auditable record of the consent history. It helps ensure accountability and assists in identifying any unauthorized or suspicious activities.

\subsection{Privacy Safeguards}

Protecting patient privacy is a top priority in healthcare systems \cite{CHENG2022342,albalwy2021blockchain}. The Ethereum-based consent management system used in the NewbornTime project includes strong privacy protections, as summarized in Table~\ref{tab:privacy-safeguards}:

\subsubsection{Encryption}

Sensitive patient information, like mother's names, national identification numbers, and phone numbers, is encrypted before being stored on the blockchain. This encryption makes sure that even if unauthorized access happens, the data stays confidential and cannot be read without the correct decryption keys.

\subsubsection{Data Minimization}

The system follows the principle of data minimization, where only the essential information needed for consent management is collected and stored. This approach lowers the risk linked to storing and handling sensitive information, reducing potential weaknesses.

\subsubsection{Decentralization}

The decentralized structure of the Ethereum blockchain makes sure that patient data is not kept in a single, vulnerable location prone to data breaches. Instead, the data is spread across multiple nodes, which improves protection against attacks and strengthens data security \cite{7990130}.

Incorporating these security measures and privacy protections together creates a strong foundation for the Ethereum-based healthcare consent management system, building data integrity, and patient confidence within the NewbornTime project.

The consent management system incorporates a comprehensive set of security measures and privacy safeguards to ensure the confidentiality, integrity, and availability of patient data. These measures are critical and maintaining compliance with healthcare data protection regulations.

Table~\ref{tab:security-considerations} summarizes the key security considerations implemented in the system. These include robust access control mechanisms, secure data storage leveraging the tamper-resistant blockchain, and transparent audit trails that log all consent-related changes. Together, these measures prevent unauthorized access, ensure data integrity, and enhance accountability.

\begin{table}[!htbp]
    \centering
    \caption{Summary of Security Considerations}
    \label{tab:security-considerations}
    \begin{tabular}{|p{3.7cm}|p{9.0cm}|}
        \hline
        \textbf{Security Aspect} & \textbf{Description} \\
        \hline \hline
        Access Control Mechanisms & Control access to critical functions through the \texttt{onlyOwner} modifier, ensuring only authorized entities can execute them. \\
        Secure Data Storage & Utilize the tamper-resistant and immutable nature of the blockchain for secure patient data storage, employing cryptographic techniques for confidentiality. \\
        Auditability and Transparency & Log consent status changes using the \texttt{ConsentChanged} event, enhancing accountability and detection of unauthorized activities. \\
        \hline
    \end{tabular}
\end{table}

Table~\ref{tab:privacy-safeguards} outlines the privacy safeguards integrated into the system. These include the encryption of sensitive patient data, adherence to the principle of data minimization, and the decentralized architecture of the Ethereum blockchain. By applying these principles, the system minimizes vulnerabilities and ensures compliance with privacy standards, protecting user data at all times.

\begin{table}[!htbp]
    \centering
    \caption{Summary of Privacy Safeguards}
    \label{tab:privacy-safeguards}
    \begin{tabular}{|p{3.7cm}|p{9.0cm}|}
        \hline
        \textbf{Privacy Aspect} & \textbf{Description} \\
        \hline \hline
        Encryption & Apply encryption to sensitive patient information before storage, ensuring confidentiality and data security. \\
        Data Minimization & Collect and store only essential information required for consent management, reducing the risk associated with sensitive data. \\
        Decentralization & Leverage the decentralized Ethereum blockchain to distribute patient data across multiple nodes, enhancing data security and resilience. \\
        \hline
    \end{tabular}
\end{table}

Personal identifiable information (PII) is not stored on the blockchain; instead, the blockchain records only consent-related metadata, ensuring privacy while maintaining transparency. Upon withdrawal of consent, a new transaction is recorded on the blockchain to indicate revocation, and the associated personal data is promptly deleted from off-chain storage systems, ensuring compliance with data protection regulations such as GDPR.

\section{Experimental Results}
Let $H$ be the set of all healthcare providers, $P$ be the set of all patients, and $\mathcal{C}$ be the set of all consent records. For each patient $p \in P$, $\mathcal{C}_p$ represents the set of consent records linked to patient $p$. Each consent record $c \in \mathcal{C}_p$ is formally defined as follows:
$$
c = (p, h, g, m, n, ph, t)
$$
where:
\begin{itemize}
    \item $p$ is the patient's Ethereum address,
    \item $h$ is the healthcare provider's Ethereum address,
    \item $g$ denotes the consent status (1 for granted, 0 for revoked),
    \item $m$ is the mother's name (a string),
    \item $n$ is the national identification number (an integer),
    \item $ph$ is the phone number (a string), and
    \item $t$ is the timestamp (a Unix timestamp).
\end{itemize}

The \texttt{HealthcareConsent.sol} contract manages consent status changes within $\mathcal{C}$. Access control is enforced by the \texttt{onlyOwner} modifier, which allows only the contract owner to call certain functions. Additionally, the contract includes an \texttt{event}, $\texttt{ConsentChanged}(p, h, g, t)$, which records consent changes, providing a permanent and clear audit trail.

\subsection{Gas Costs}
The gas costs \cite{8038517} associated with various operations of the \texttt{HealthcareConsent.sol} smart contract were measured. Gas costs represent the computational resources required to execute operations on the Ethereum blockchain. Each operation, such as adding, querying, or revoking consent, consumes a specific amount of gas, which translates to transaction fees paid to miners for processing and securing these actions. Table~\ref{tab:gas-costs} shows the gas used for adding, querying, and revoking consent records.

\begin{table}[htbp]
    \centering
    \caption{Gas Costs for Different Operations}
    \label{tab:gas-costs}
    \begin{tabular}{|l|r|r|}
        \hline
        \textbf{Operation} & \textbf{Gas Used} & \textbf{Execution Time (ms)} \\
        \hline \hline
        \multirow{5}{*}{\textbf{Add Consent}} 
            & 175719 & 155.850887 \\
            & 160719 & 129.161119 \\
            & 160719 & 136.569977 \\
            & 160719 & 169.187069 \\
            & 160719 & 133.695841 \\ \hline \hline
        \multirow{1}{*}{\textbf{Query Consent}} 
            & 0 & 105.110168 \\ \hline \hline
        \multirow{5}{*}{\textbf{Revoke Consent}} 
            & 37035 & 69.026947 \\
            & 41601 & 97.961903 \\
            & 46167 & 74.846983 \\
            & 50733 & 73.647261 \\
            & 55299 & 72.370768 \\
        \hline
    \end{tabular}
\end{table}

The gas cost for adding a consent record was observed to be significantly higher than querying or revoking a consent. Query operations do not require gas as they are read-only, while revoking a consent incurs a cost due to state modification.

\subsection{Transaction Throughput}
\begin{figure}[H]
    \centering
    \includegraphics[width=0.8\textwidth]{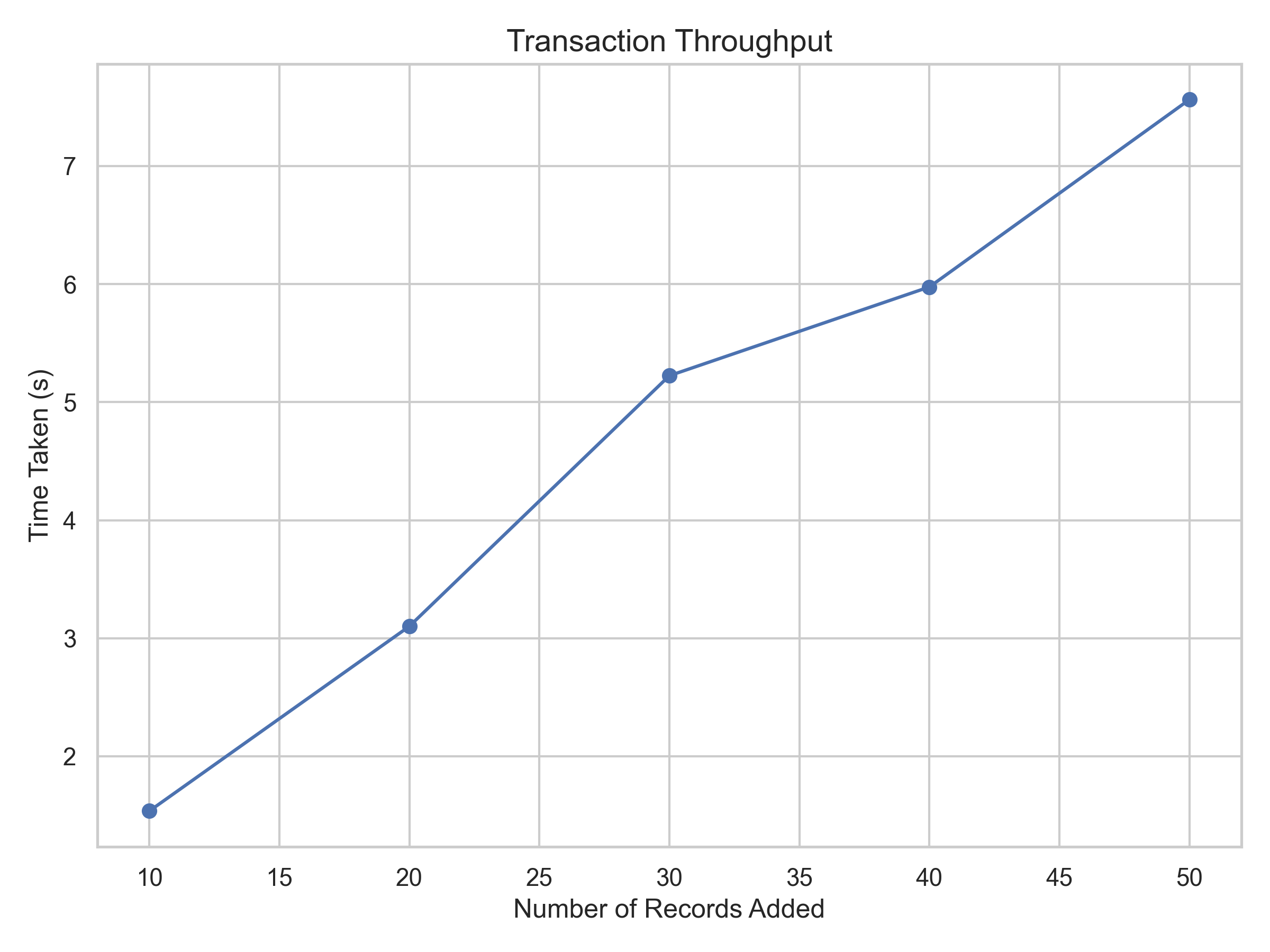}
    \caption{Transaction Throughput Over Time}
    \label{fig:throughput}
\end{figure}

To assess the system's ability to handle multiple transactions, we measured the time taken to add batches of consent records. Figure~\ref{fig:throughput} shows the transaction throughput as the number of records increases.

The system was able to handle up to 50 records within a reasonable time frame, with a noticeable increase in transaction time as the batch size increased. This suggests the system is capable of handling moderate loads but may require optimization for larger-scale operations.

\subsection{Scalability Analysis}
To test the scalability of the system, we measured the gas used and time taken to add increasingly larger batches of consent records. Table~\ref{tab:scalability} summarizes the scalability results.

\begin{table}[htbp]
    \centering
    \caption{Scalability Test Results}
    \label{tab:scalability}
    \begin{tabular}{|l|r|r|}
        \hline
        \textbf{Records Added} & \textbf{Gas Used} & \textbf{Time Taken (ms)} \\
        \hline \hline
10 & 1606870 & 183.220911 \\
50 & 8034350 & 182.528577 \\
100 & 16068700 & 192.539830 \\
500 & 80343476 & 207.127324 \\
        \hline
    \end{tabular}
\end{table}

The results indicate that gas consumption and time scale with the number of records. This behavior is expected due to the immutable and decentralized nature of the Ethereum blockchain.

\subsection{Privacy and Data Minimization}
To evaluate the impact of data minimization on privacy and system performance, we compared the gas costs of storing full consent records versus minimal consent records. Table~\ref{tab:privacy} summarizes the gas usage and execution time for these two types of records.

The results indicate that storing minimal consent data significantly reduces gas costs (102,437 gas units compared to 160,747 gas units for full data) and execution time (144.618 ms versus 164.344 ms). These findings highlight the efficiency benefits of storing only essential information, such as anonymized or minimal data, while supporting the principle of data minimization.

\begin{table}[htbp]
    \centering
    \caption{Gas Costs for Minimal vs Full Consent Records}
    \label{tab:privacy}
    \begin{tabular}{|l|r|r|}
        \hline
        \textbf{Data Type} & \textbf{Gas Used} & \textbf{Execution Time (ms)} \\
        \hline \hline
        Full Data & 160747 & 164.344788 \\
        Minimal Data & 102437 & 144.618034 \\
        \hline
    \end{tabular}
\end{table}

The results show that storing minimal consent data significantly reduces gas costs, which supports the principle of data minimization. This ensures that only necessary data is stored on-chain, reducing storage overhead and enhancing privacy.

\section{Discussion}

The experimental results demonstrate the effectiveness of the Ethereum-based healthcare consent management system in balancing security, privacy, and scalability, with certain limitations. In this section, we discuss the key findings and their broader implications for healthcare and blockchain-based consent management systems.

\subsection{Data Privacy and Security}
The system’s privacy safeguards, particularly encryption and data minimization, were validated through the experiments. Encryption ensures that even if unauthorized access occurs, sensitive patient information such as national IDs and phone numbers remains confidential. The principle of data minimization reduces the amount of sensitive data stored on-chain, thus decreasing the risk of data breaches.

The decentralized nature of the Ethereum network ensures that patient data is not stored in a single centralized repository, reducing the risk of breaches and enhancing overall data security.

The system ensures that personal data is not stored on the blockchain, minimizing privacy risks. The consent withdrawal mechanism triggers immediate deletion of associated data from off-chain storage, demonstrating the system's commitment to data protection. In the context of the NewbornTime project, the system ensures that sensitive video data is only accessed and processed with valid participant consent, thereby protecting participant privacy and maintaining research integrity.

\subsection{Scalability and Performance}
The scalability experiments revealed that the system performs well under moderate loads but faces challenges when scaling to handle large volumes of consent records \cite{10.1007/978-3-662-58387-6_24}. As shown in Table~\ref{tab:scalability-results}, gas costs increase with the number of consent records, presenting a limitation for large-scale healthcare environments.

The transaction throughput remains sufficient for moderate-scale healthcare settings, but for larger-scale systems, optimization strategies such as Layer 2 scaling solutions (e.g., rollups) could be employed to improve scalability and reduce transaction costs \cite{10.1145/3243734.3243853}.

Table \ref{tab:scalability-results} shows the summary of experimental results.

\begin{table}[htbp]
    \centering
    \caption{Summary of Experimental Results}
    \label{tab:scalability-results}
    \begin{tabular}{|l|p{7cm}|}
        \hline
        \textbf{Aspect} & \textbf{Key Findings} \\
        \hline \hline
        \textbf{Transparency} & Blockchain immutability and audit trails enhance trust between patients and healthcare providers. \\
        \hline
        \textbf{Data Privacy and Security} & Encryption and data minimization ensure confidentiality of sensitive patient information, while decentralization improves data security. \\
        \hline
        \textbf{Scalability} & System performs well under moderate loads; however, gas costs and time increase linearly with the number of records, limiting scalability. \\
        \hline
        \textbf{Trade-offs} & High gas costs and slower transaction times in large-scale settings. Optimization needed for scalability. \\
        \hline
    \end{tabular}
\end{table}

\subsection{Trade-offs and Future Work}
While the system successfully addresses the key challenges of security, and privacy, there are trade-offs between security, scalability, and cost. The decentralized nature of the blockchain provides security but results in higher gas costs and slower transaction times, particularly as the system scales.

Future work should focus on optimizing gas usage, exploring alternative blockchain platforms, or integrating more efficient consensus mechanisms. Moreover, the development of user-friendly interfaces for patients and healthcare providers would enhance usability and adoption in real-world healthcare settings.

Future research should include user studies to assess the level of trust and acceptance among participants, especially mothers-to-be, to ensure the system meets their needs and expectations \cite{info:doi/10.2196/50339}. While the current focus is on research consent management, future work could explore adapting the system for clinical healthcare applications, addressing the specific needs and regulations of patient care environments\cite{electronics9010094}. A limitation of the current implementation is the low usage of the digital portal for consent management, with participants preferring direct contact with research staff. Future improvements could focus on enhancing user education, simplifying the interface, and providing targeted training to increase engagement with the digital platform \cite{velmovitsky2020blockchain}.

\section{Conclusion}

This study has presented a complete consent management system developed for the NewbornTime project, which focuses on solving key challenges related to patient consent, and data security in healthcare. The system is built around the \texttt{HealthcareConsent.sol} smart contract on the Ethereum blockchain, which ensures transparency, data integrity, and privacy when handling patient consent.

The modular design of the system, shown in the system overview figure, includes the Consent Portal, the Cloud System for Blockchain, the Cloud System for Reporting, and real-time Consent Verification in the delivery room. This setup allows mothers to provide consent securely and enables healthcare professionals to verify consent and generate a Study ID that links the mother's and newborn's data for research purposes.

A major part of the system is the \texttt{Consent Statistics Dashboard}, which gives healthcare providers and researchers with clear and useful information about consent trends, status, and records. This feature improves the management of consent processes and supports better decision-making through real-time data analysis.

To protect sensitive data, the system uses strong security and privacy measures, such as data encryption, access control, and tamper-proof storage of the blockchain. These measures ensure that only authorized users can access patient data and that the system meets data protection standards like GDPR.

The system has shown promising performance in managing consent records, but some challenges remain. High gas costs and scalability issues with Ethereum need to be addressed. Future work will focus on making the system more efficient and scalable, possibly by exploring alternative blockchain solutions or optimization methods.

The proposed system offers a secure and reliable solution for managing patient consent in healthcare. It combines blockchain technology, a user-friendly dashboard, and secure data handling to improve efficiency. Future developments will aim to expand the system’s use in other healthcare and research projects while improving performance and lowering costs.

\bibliographystyle{elsarticle-num-names} 
\bibliography{ref}

\begin{thebibliography}{22}
\expandafter\ifx\csname natexlab\endcsname\relax\def\natexlab#1{#1}\fi
\providecommand{\url}[1]{\texttt{#1}}
\providecommand{\href}[2]{#2}
\providecommand{\path}[1]{#1}
\providecommand{\DOIprefix}{doi:}
\providecommand{\ArXivprefix}{arXiv:}
\providecommand{\URLprefix}{URL: }
\providecommand{\Pubmedprefix}{pmid:}
\providecommand{\doi}[1]{\href{http://dx.doi.org/#1}{\path{#1}}}
\providecommand{\Pubmed}[1]{\href{pmid:#1}{\path{#1}}}
\providecommand{\bibinfo}[2]{#2}
\ifx\xfnm\relax \def\xfnm[#1]{\unskip,\space#1}\fi
\bibitem[{Griggs et~al.(2018)Griggs, Ossipova, Kohlios, Baccarini, Howson, and
  Hayajneh}]{griggs2018healthcare}
\bibinfo{author}{K.~N. Griggs}, \bibinfo{author}{O.~Ossipova},
  \bibinfo{author}{C.~P. Kohlios}, \bibinfo{author}{A.~N. Baccarini},
  \bibinfo{author}{E.~A. Howson}, \bibinfo{author}{T.~Hayajneh},
\newblock \bibinfo{title}{Healthcare blockchain system using smart contracts
  for secure automated remote patient monitoring},
\newblock \bibinfo{journal}{Journal of medical systems} \bibinfo{volume}{42}
  (\bibinfo{year}{2018}) \bibinfo{pages}{1--7}.
\bibitem[{Shah et~al.(2024)Shah, Thornton, Turrin et~al.}]{shah2024informed}
\bibinfo{author}{P.~Shah}, \bibinfo{author}{I.~Thornton},
  \bibinfo{author}{D.~Turrin}, et~al.,
\newblock \bibinfo{title}{Informed consent},
\newblock \bibinfo{journal}{StatPearls}  (\bibinfo{year}{2024}).
\bibitem[{Obaidi et~al.(2024)Obaidi, Elkhyatt, Alzubaidi, and
  Househ}]{obaidi2024econsent}
\bibinfo{author}{H.~Obaidi}, \bibinfo{author}{Y.~Elkhyatt},
  \bibinfo{author}{M.~Alzubaidi}, \bibinfo{author}{M.~Househ},
\newblock \bibinfo{title}{Use of e-consent in healthcare settings: A scoping
  review},
\newblock \bibinfo{journal}{Stud Health Technol Inform} \bibinfo{volume}{316}
  (\bibinfo{year}{2024}) \bibinfo{pages}{1064--1068}.
\bibitem[{Zheng et~al.(2020)Zheng, Xie, Dai et~al.}]{zheng2020smartcontracts}
\bibinfo{author}{Z.~Zheng}, \bibinfo{author}{S.~Xie}, \bibinfo{author}{H.~Dai},
  et~al.,
\newblock \bibinfo{title}{An overview on smart contracts: Challenges, advances,
  and platforms},
\newblock \bibinfo{journal}{Future Generation Computer Systems}
  \bibinfo{volume}{105} (\bibinfo{year}{2020}) \bibinfo{pages}{475--491}.
\bibitem[{Engan et~al.(2023)Engan, Rettedal et~al.}]{engan2023newborntime}
\bibinfo{author}{K.~Engan}, \bibinfo{author}{S.~I. Rettedal}, et~al.,
\newblock \bibinfo{title}{Newborn time - improved newborn care based on video
  and artificial intelligence - study protocol},
\newblock \bibinfo{journal}{BMC Digital Health} \bibinfo{volume}{1}
  (\bibinfo{year}{2023}) \bibinfo{pages}{10}.
\bibitem[{Bjorland et~al.(2019)Bjorland, {\O}ymar, Ersdal, and
  Rettedal}]{bjorland2019incidence}
\bibinfo{author}{P.~A. Bjorland}, \bibinfo{author}{K.~{\O}ymar},
  \bibinfo{author}{H.~L. Ersdal}, \bibinfo{author}{S.~I. Rettedal},
\newblock \bibinfo{title}{Incidence of newborn resuscitative interventions at
  birth and short-term outcomes: a regional population-based study},
\newblock \bibinfo{journal}{BMJ paediatrics open} \bibinfo{volume}{3}
  (\bibinfo{year}{2019}).
\bibitem[{Wyckoff et~al.(2020)Wyckoff, Wyllie, Aziz, de~Almeida, Fabres, Fawke,
  Guinsburg, Hosono, Isayama, Kapadia et~al.}]{wyckoff2020neonatal}
\bibinfo{author}{M.~H. Wyckoff}, \bibinfo{author}{J.~Wyllie},
  \bibinfo{author}{K.~Aziz}, \bibinfo{author}{M.~F. de~Almeida},
  \bibinfo{author}{J.~Fabres}, \bibinfo{author}{J.~Fawke},
  \bibinfo{author}{R.~Guinsburg}, \bibinfo{author}{S.~Hosono},
  \bibinfo{author}{T.~Isayama}, \bibinfo{author}{V.~S. Kapadia}, et~al.,
\newblock \bibinfo{title}{Neonatal life support: 2020 international consensus
  on cardiopulmonary resuscitation and emergency cardiovascular care science
  with treatment recommendations},
\newblock \bibinfo{journal}{Circulation} \bibinfo{volume}{142}
  (\bibinfo{year}{2020}) \bibinfo{pages}{S185--S221}.
\bibitem[{Garc{\'\i}a-Torres et~al.(2022)Garc{\'\i}a-Torres, Meinich-Bache,
  Brunner, Johannessen, Rettedal, and Engan}]{garcia2022towards}
\bibinfo{author}{J.~Garc{\'\i}a-Torres}, \bibinfo{author}{{\O}.~Meinich-Bache},
  \bibinfo{author}{S.~Brunner}, \bibinfo{author}{A.~Johannessen},
  \bibinfo{author}{S.~Rettedal}, \bibinfo{author}{K.~Engan},
\newblock \bibinfo{title}{Towards using thermal cameras in birth detection},
\newblock in: \bibinfo{booktitle}{2022 IEEE 14th Image, Video, and
  Multidimensional Signal Processing Workshop (IVMSP)},
  \bibinfo{organization}{IEEE}, \bibinfo{year}{2022}, pp.
  \bibinfo{pages}{1--5}.
\bibitem[{Garc{\'\i}a-Torres et~al.(2024)Garc{\'\i}a-Torres, Meinich-Bache,
  Rettedal, Kibsgaard, Brunner, and Engan}]{garcia2024comparative}
\bibinfo{author}{J.~Garc{\'\i}a-Torres}, \bibinfo{author}{{\O}.~Meinich-Bache},
  \bibinfo{author}{S.~I. Rettedal}, \bibinfo{author}{A.~Kibsgaard},
  \bibinfo{author}{S.~Brunner}, \bibinfo{author}{K.~Engan},
\newblock \bibinfo{title}{Comparative analysis of binary and multiclass
  activity recognition in high-quality newborn resuscitation videos},
\newblock in: \bibinfo{booktitle}{Northern Lights Deep Learning Conference
  2024}, \bibinfo{year}{2024}.
\bibitem[{Kolstad et~al.(2024)Kolstad, Garc{\'\i}a-Torres, Brunner,
  Johannessen, Foglia, Ersdal, Meinich-Bache, and
  Rettedal}]{kolstad2024detection}
\bibinfo{author}{V.~Kolstad}, \bibinfo{author}{J.~Garc{\'\i}a-Torres},
  \bibinfo{author}{S.~Brunner}, \bibinfo{author}{A.~Johannessen},
  \bibinfo{author}{E.~Foglia}, \bibinfo{author}{H.~Ersdal},
  \bibinfo{author}{{\O}.~Meinich-Bache}, \bibinfo{author}{S.~Rettedal},
\newblock \bibinfo{title}{Detection of time of birth and cord clamping using
  thermal video in the delivery room},
\newblock \bibinfo{journal}{Frontiers in Pediatrics} \bibinfo{volume}{12}
  (\bibinfo{year}{2024}) \bibinfo{pages}{1342415}.
\bibitem[{Szabo(1997)}]{szabo1997formalizing}
\bibinfo{author}{N.~Szabo},
\newblock \bibinfo{title}{Formalizing and securing relationships on public
  networks},
\newblock \bibinfo{journal}{First monday}  (\bibinfo{year}{1997}).
\bibitem[{Buterin et~al.(2014)}]{buterin2014next}
\bibinfo{author}{V.~Buterin}, et~al.,
\newblock \bibinfo{title}{A next-generation smart contract and decentralized
  application platform},
\newblock \bibinfo{journal}{white paper} \bibinfo{volume}{3}
  (\bibinfo{year}{2014}) \bibinfo{pages}{2--1}.
\bibitem[{Albanese et~al.(2020)Albanese, Calbimonte, Schumacher, and
  Calvaresi}]{albanese2020dynamic}
\bibinfo{author}{G.~Albanese}, \bibinfo{author}{J.-P. Calbimonte},
  \bibinfo{author}{M.~Schumacher}, \bibinfo{author}{D.~Calvaresi},
\newblock \bibinfo{title}{Dynamic consent management for clinical trials via
  private blockchain technology},
\newblock \bibinfo{journal}{Journal of ambient intelligence and humanized
  computing} \bibinfo{volume}{11} (\bibinfo{year}{2020})
  \bibinfo{pages}{4909--4926}.
\bibitem[{Cheng et~al.(2022)Cheng, Yan, and Wang}]{CHENG2022342}
\bibinfo{author}{C.~Cheng}, \bibinfo{author}{B.~Yan},
  \bibinfo{author}{G.~Wang},
\newblock \bibinfo{title}{The blockchain based access control scheme for the
  internet of things},
\newblock \bibinfo{journal}{Procedia Computer Science} \bibinfo{volume}{202}
  (\bibinfo{year}{2022}) \bibinfo{pages}{342--347}. \URLprefix
  \url{https://www.sciencedirect.com/science/article/pii/S1877050922005804}.
  \DOIprefix\doi{https://doi.org/10.1016/j.procs.2022.04.046},
  \bibinfo{note}{international Conference on Identification, Information and
  Knowledge in the internet of Things, 2021}.
\bibitem[{Albalwy et~al.(2021)Albalwy, Brass, Davies
  et~al.}]{albalwy2021blockchain}
\bibinfo{author}{F.~Albalwy}, \bibinfo{author}{A.~Brass},
  \bibinfo{author}{A.~Davies}, et~al.,
\newblock \bibinfo{title}{A blockchain-based dynamic consent architecture to
  support clinical genomic data sharing (consentchain): Proof-of-concept
  study},
\newblock \bibinfo{journal}{JMIR medical informatics} \bibinfo{volume}{9}
  (\bibinfo{year}{2021}) \bibinfo{pages}{e27816}.
\bibitem[{Xia et~al.(2017)Xia, Sifah, Asamoah, Gao, Du, and Guizani}]{7990130}
\bibinfo{author}{Q.~Xia}, \bibinfo{author}{E.~B. Sifah}, \bibinfo{author}{K.~O.
  Asamoah}, \bibinfo{author}{J.~Gao}, \bibinfo{author}{X.~Du},
  \bibinfo{author}{M.~Guizani},
\newblock \bibinfo{title}{Medshare: Trust-less medical data sharing among cloud
  service providers via blockchain},
\newblock \bibinfo{journal}{IEEE Access} \bibinfo{volume}{5}
  (\bibinfo{year}{2017}) \bibinfo{pages}{14757--14767}.
  \DOIprefix\doi{10.1109/ACCESS.2017.2730843}.
\bibitem[{Pongnumkul et~al.(2017)Pongnumkul, Siripanpornchana, and
  Thajchayapong}]{8038517}
\bibinfo{author}{S.~Pongnumkul}, \bibinfo{author}{C.~Siripanpornchana},
  \bibinfo{author}{S.~Thajchayapong},
\newblock \bibinfo{title}{Performance analysis of private blockchain platforms
  in varying workloads},
\newblock in: \bibinfo{booktitle}{2017 26th International Conference on
  Computer Communication and Networks (ICCCN)}, \bibinfo{year}{2017}, pp.
  \bibinfo{pages}{1--6}. \DOIprefix\doi{10.1109/ICCCN.2017.8038517}.
\bibitem[{Gencer et~al.(2018)Gencer, Basu, Eyal, van Renesse, and
  Sirer}]{10.1007/978-3-662-58387-6_24}
\bibinfo{author}{A.~E. Gencer}, \bibinfo{author}{S.~Basu},
  \bibinfo{author}{I.~Eyal}, \bibinfo{author}{R.~van Renesse},
  \bibinfo{author}{E.~G. Sirer},
\newblock \bibinfo{title}{Decentralization in bitcoin and ethereum networks},
\newblock in: \bibinfo{booktitle}{Financial Cryptography and Data Security:
  22nd International Conference, FC 2018, Nieuwpoort, Cura\c{c}ao, February 26
  – March 2, 2018, Revised Selected Papers},
  \bibinfo{publisher}{Springer-Verlag}, \bibinfo{address}{Berlin, Heidelberg},
  \bibinfo{year}{2018}, p. \bibinfo{pages}{439–457}. \URLprefix
  \url{https://doi.org/10.1007/978-3-662-58387-6_24}.
  \DOIprefix\doi{10.1007/978-3-662-58387-6_24}.
\bibitem[{Zamani et~al.(2018)Zamani, Movahedi, and
  Raykova}]{10.1145/3243734.3243853}
\bibinfo{author}{M.~Zamani}, \bibinfo{author}{M.~Movahedi},
  \bibinfo{author}{M.~Raykova},
\newblock \bibinfo{title}{Rapidchain: Scaling blockchain via full sharding},
\newblock in: \bibinfo{booktitle}{Proceedings of the 2018 ACM SIGSAC Conference
  on Computer and Communications Security}, CCS '18,
  \bibinfo{publisher}{Association for Computing Machinery},
  \bibinfo{address}{New York, NY, USA}, \bibinfo{year}{2018}, p.
  \bibinfo{pages}{931–948}. \URLprefix
  \url{https://doi.org/10.1145/3243734.3243853}.
  \DOIprefix\doi{10.1145/3243734.3243853}.
\bibitem[{Charles et~al.(2024)Charles, van~der Waal, Flach, Bisschop, van~der
  Waal, Es-Sbai, and McLeod}]{info:doi/10.2196/50339}
\bibinfo{author}{W.~M. Charles}, \bibinfo{author}{M.~B. van~der Waal},
  \bibinfo{author}{J.~Flach}, \bibinfo{author}{A.~Bisschop},
  \bibinfo{author}{R.~X. van~der Waal}, \bibinfo{author}{H.~Es-Sbai},
  \bibinfo{author}{C.~J. McLeod},
\newblock \bibinfo{title}{Blockchain-based dynamic consent and its applications
  for patient-centric research and health information sharing: Protocol for an
  integrative review},
\newblock \bibinfo{journal}{JMIR Res Protoc} \bibinfo{volume}{13}
  (\bibinfo{year}{2024}) \bibinfo{pages}{e50339}. \URLprefix
  \url{https://www.researchprotocols.org/2024/1/e50339}.
  \DOIprefix\doi{10.2196/50339}.
\bibitem[{Khatoon(2020)}]{electronics9010094}
\bibinfo{author}{A.~Khatoon},
\newblock \bibinfo{title}{A blockchain-based smart contract system for
  healthcare management},
\newblock \bibinfo{journal}{Electronics} \bibinfo{volume}{9}
  (\bibinfo{year}{2020}). \URLprefix
  \url{https://www.mdpi.com/2079-9292/9/1/94}.
  \DOIprefix\doi{10.3390/electronics9010094}.
\bibitem[{Velmovitsky et~al.(2020)Velmovitsky, Miranda, Vaillancourt, Donovska,
  Teague, and Morita}]{velmovitsky2020blockchain}
\bibinfo{author}{P.~E. Velmovitsky}, \bibinfo{author}{P.~A. D. S. E.~S.
  Miranda}, \bibinfo{author}{H.~Vaillancourt}, \bibinfo{author}{T.~Donovska},
  \bibinfo{author}{J.~Teague}, \bibinfo{author}{P.~P. Morita},
\newblock \bibinfo{title}{A blockchain-based consent platform for active
  assisted living: modeling study and conceptual framework},
\newblock \bibinfo{journal}{Journal of medical Internet research}
  \bibinfo{volume}{22} (\bibinfo{year}{2020}) \bibinfo{pages}{e20832}.

\end{thebibliography}





\end{document}